\documentclass[twocolumn,preprintnumbers,unsortedaddress,amsmath,prb,nobibnotes,altaffilletter]{revtex4}

\usepackage{graphics,graphicx}

\preprint{Physics of Fluids}

\begin{document}

\title{Fluctuations of statistics among subregions of a turbulence velocity field}

\author{Hideaki Mouri}
\email{hmouri@mri-jma.go.jp}
\author{Akihiro Hori}
\altaffiliation[Also at ]{Meteorological and Environmental Sensing Technology, Inc., Nanpeidai, Ami 300-0312, Japan}
\affiliation{Meteorological Research Institute, Nagamine, Tsukuba 305-0052, Japan}

\author{Masanori Takaoka}
\email{mtakaoka@mail.doshisha.ac.jp}
\affiliation{Department of Mechanical Engineering, Doshisha University, Kyotanabe, Kyoto 610-0321, Japan}



\begin{abstract}
To study subregions of a turbulence velocity field, a long record of velocity data of grid turbulence is divided into smaller segments. For each segment, we calculate statistics such as the mean rate of energy dissipation and the mean energy at each scale. Their values significantly fluctuate, in lognormal distributions at least as a good approximation. Each segment is not under equilibrium between the mean rate of energy dissipation and the mean rate of energy transfer that determines the mean energy. These two rates still correlate among segments when their length exceeds the correlation length. Also between the mean rate of energy dissipation and the mean total energy, there is a correlation characterized by the Reynolds number for the whole record, implying that the large-scale flow affects each of the segments.
\end{abstract}

\maketitle

\section{Introduction}
\label{s1}

For locally isotropic turbulence, Kolmogorov \cite{k41} considered that small-scale statistics are uniquely determined by the kinematic viscosity $\nu$ and the mean rate of energy dissipation $\langle \varepsilon \rangle$. The Kolmogorov velocity $u_{\rm K} = (\nu \langle \varepsilon \rangle)^{1/4}$ and the Kolmogorov length $\eta = (\nu ^3 / \langle \varepsilon \rangle)^{1/4}$ determine the statistics of velocity increment $\delta u_r = u(x+r)-u(x)$ at scale $r$ as
\begin{equation}
\frac{\langle \delta u_r^n \rangle}{u_{\rm K}^n} = F_n \left( \frac{r}{\eta} \right)
\quad
\mbox{for} \ \ n=2,3,4,....
\end{equation}
Here $\langle \cdot \rangle$ denotes an average over position $x$, and $F_n$ is a universal function. The universality is known to hold well. While $\langle \delta u_r^n \rangle$ at each $r$ is different in different velocity fields, $\langle \varepsilon \rangle$ and hence $u_{\rm K}^n$ and $\eta$ are accordingly different. That is, $\langle \varepsilon \rangle$ is in equilibrium with the mean rate of energy transfer that determines $\langle \delta u_r^n \rangle$.

However, the universality of small-scale statistics might not be exact. To argue against the exact universality, Landau\cite{ll59} pointed out that the local rate of energy dissipation $\varepsilon$ fluctuates over large scales. This fluctuation is not universal and is always significant.\cite{po97,c03,mouri06} In fact, the large-scale flow or the configuration for turbulence production appears to affect some small-scale statistics.\cite{mouri06,k92,pgkz93,ss96,sd98,mininni06}

Obukhov\cite{o62} discussed that Kolmogorov's theory\cite{k41} still holds in an ensemble of ``pure'' subregions where $\varepsilon$ is constant at a certain value. Then, the $\varepsilon$ value represents the rate of energy transfer averaged over those subregions. For the whole region, small-scale statistics reflect the large-scale flow through the large-scale fluctuation of the $\varepsilon$ value. The idea that turbulence consists of some elementary subregions is of interest even now.\cite{ss96} We study statistics among subregions in terms of the effect of large scales on small scales, by using a long record of velocity data obtained in grid turbulence.

\section{Experiment}
\label{s2}

The experiment was done in a wind tunnel of the Meteorological Research Institute. Its test section had the size of 18, 3, and 2\,m in the streamwise, spanwise, and floor-normal directions. We placed a grid across the entrance to the test section. The grid consisted of two layers of uniformly spaced rods, with axes in the two layers at right angles. The cross section of the rods was $0.04 \times 0.04$\,m$^2$. The separation of the axes of adjacent rods was 0.20\,m.

On the tunnel axis at 4\,m downstream of the grid, we simultaneously measured the streamwise ($U+u$) and spanwise ($v$) velocities. Here $U$ is the average while $u(t)$ and $v(t)$ are fluctuations as a function of time $t$. We used a hot-wire anemometer with a crossed-wire probe. The wires were made of platinum-plated tungsten, 5\,$\mu$m in diameter, 1.25\,mm in sensing length, 1\,mm in separation, oriented at $\pm 45^{\circ}$ to the streamwise direction, and 280\,$^{\circ}$C in temperature. The signal was linearized, low-pass filtered at 35\,kHz, and then digitally sampled at $f_s = 70$\,kHz. We obtained as long as $4 \times 10^8$ data.

The calibration coefficient, with which the flow velocity is proportional to the anemometer signal, depends on the condition of the hot wires and thereby varied slowly in time. We determine the coefficient so as to have $U = 21.16$\,m\,s$^{-1}$ for each segment with $4 \times 10^6$ data. Within each segment, the coefficient varied by $\pm 0.4$\% at most. Also varied slowly in time the flow temperature and hence the kinematic viscosity $\nu$. We adopt $\nu = 1.42 \times 10^{-5}$\,m$^2$\,s$^{-1}$ based on the mean flow temperature, 11.8$^{\circ}$C. The temperature variation, $\pm1.2$\,$^{\circ}$C, corresponds to the $\nu$ variation of $\pm 0.7$\%. These variations are small and ignored here.

Taylor's frozen-eddy hypothesis, i.e., $x = -Ut$, is used to obtain $u(x)$ and $v(x)$ from $u(t)$ and $v(t)$. This hypothesis requires a small value of $\langle u^2 \rangle^{1/2}/U$. The value in our experiment, 0.05, is small enough. Since $u(t)$ and $v(t)$ are stationary, $u(x)$ and $v(x)$ are homogeneous, although grid turbulence decays along the streamwise direction in the wind tunnel. We are mostly interested in scales up to about the typical scale for energy-containing eddies, which is much less than the tunnel size. Over such scales, fluctuations of $u(x)$ and $v(x)$ correspond to spatial fluctuations that were actually present in the wind tunnel.\cite{note0} Those over the larger scales do not. They have to be interpreted as fluctuations over long timescales described in terms of large length scales.\cite{note1}

\begingroup
\squeezetable
\begin{table}[t]
\caption{\label{t1} Turbulence parameters: mean energy dissipation rate $\langle \varepsilon \rangle$, rms velocity fluctuations $\langle u^2 \rangle ^{1/2}$ and $\langle v^2 \rangle ^{1/2}$, Kolmogorov velocity $u_{\rm K}$, flatness factors $\langle u^4 \rangle / \langle u^2 \rangle ^2$ and $\langle v^4 \rangle / \langle v^2 \rangle ^2$, correlation lengths $L_u$, $L_v$, and $L_{\varepsilon}$, Taylor microscale $\lambda$, Kolmogorov length $\eta$, and microscale Reynolds number Re$_{\lambda}$.}

\begin{ruledtabular}
\begin{tabular}{ll}
Quantity                                                                         &  Value                 \\ \hline               
$\langle \varepsilon \rangle = 15 \nu \langle (\partial _x v)^2 \rangle /2$      &  7.98\,m$^2$\,s$^{-3}$ \\
$\langle u^2 \rangle ^{1/2}$                                                     &  1.10\,m\,s$^{-1}$     \\
$\langle v^2 \rangle ^{1/2}$                                                     &  1.06\,m\,s$^{-1}$     \\
$u_{\rm K}  = (\nu \langle \varepsilon \rangle)^{1/4} $                          & 0.103\,m\,s$^{-1}$     \\
$\langle u^4 \rangle / \langle u^2 \rangle ^2$                                   &  3.02                  \\ 
$\langle v^4 \rangle / \langle v^2 \rangle ^2$                                   &  3.00                  \\ 
$L_u = \int^{\infty}_{0} \langle u(x+r) u(x) \rangle dr / \langle u^2 \rangle$   &  17.9\,cm              \\
$L_v = \int^{\infty}_{0} \langle v(x+r) v(x) \rangle dr / \langle v^2 \rangle$   &  4.69\,cm              \\
$L_{\varepsilon} = \int^{\infty}_{0} \langle \varepsilon (x+r) \varepsilon(x) - \langle \varepsilon \rangle ^2  \rangle dr / \langle \varepsilon ^2 - \langle \varepsilon \rangle ^2 \rangle$                                        &  0.469\,cm          \\
$\lambda = [ 2 \langle v^2 \rangle / \langle (\partial _x v )^2 \rangle ]^{1/2}$ & 0.548\,cm              \\
$\eta = (\nu ^3 / \langle \varepsilon \rangle )^{1/4}$                           & 0.0138\,cm             \\
Re$_{\lambda} =  \langle v^2 \rangle ^{1/2} \lambda / \nu$                       & 409                    \\
\end{tabular}
\end{ruledtabular}
\end{table}
\endgroup
\begin{figure}[b]
\resizebox{8cm}{!}{\includegraphics*[6cm,15.7cm][16cm,27cm]{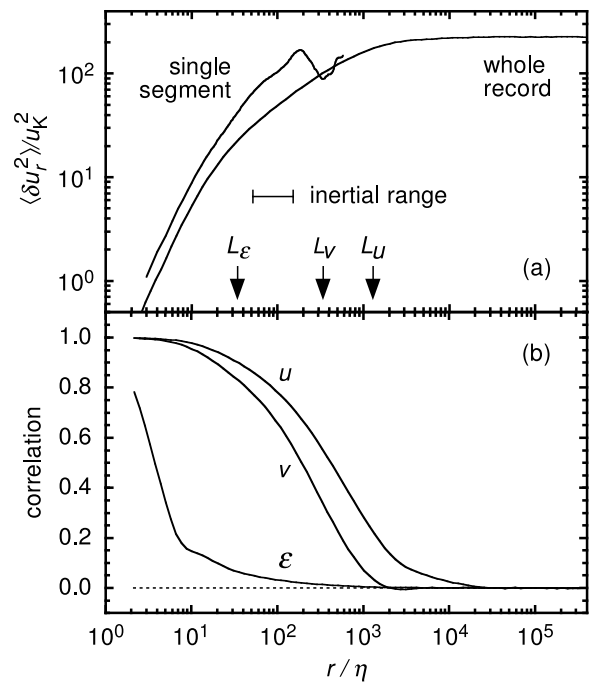}}
\caption{\label{f1} (a) $\langle \delta u_r^2 \rangle/u_{\rm K}^2$ as a function of $r/\eta$ and $\delta u_{r,R}^2/u_{{\rm K},R}^2$ for a segment with $R=10^3 \eta$ as a function of $r/\eta_R$. We indicate $L_u$, $L_v$, $L_{\varepsilon}$, and the inertial range. (b) $\langle u(x+r)u(x) \rangle /\langle u^2 \rangle$, $\langle v(x+r)v(x) \rangle /\langle v^2 \rangle$, and $\langle \varepsilon (x+r) \varepsilon(x) - \langle \varepsilon \rangle ^2  \rangle / \langle \varepsilon ^2 - \langle \varepsilon \rangle ^2 \rangle$ as a function of $r/\eta$. }
\end{figure}

Turbulence parameters are listed in Table \ref{t1}. Here and hereafter, $\langle \cdot \rangle$ is used to denote an average over the whole record. The derivative was obtained as $\partial _x v = [ 8 v(x+ \delta x)- 8 v(x- \delta x)-v(x+ 2 \delta x)+v(x- 2 \delta x)]/ 12 \delta x$ with $\delta x = U/f_s$. The local rate of energy dissipation was obtained as $\varepsilon = 15 \nu (\partial _x v)^2 /2$ instead of usual $15 \nu (\partial _x u)^2$, in order to avoid possible spurious correlations with $\delta u_r$ over small $r$ for analyses in the next section.

Figure \ref{f1} shows $\langle \delta u_r^2 \rangle / u_{\rm K}^2$, the $u$, $v$, and $\varepsilon$ correlations, and also the correlation lengths $L_u$, $L_v$, and $L_{\varepsilon}$. We see the inertial range, albeit narrow, where $\langle \delta u_r^2 \rangle$ roughly scales with $r^{2/3}$. The $u$ and $v$ correlations are significant up to $r \simeq 10^4 \eta$, which corresponds to the scale of largest eddies. The correlation length $L_u$ corresponds to the typical scale for energy-containing eddies. Since $\varepsilon$ belongs to small scales, its correlation decays quickly.

\section{Results and discussion}
\label{s3}

The data record is now divided into segments with length $R$. They correspond to subregions considered by Obukhov.\cite{o62} For each segment, we have statistics such as
\begin{eqnarray}
&&(\partial _x v)^2_R(x) = \frac{1}{R}   \int ^{x+R/2}_{x-R/2}   \left[ \frac{\partial v(x')}{\partial x'} \right]^2 dx',  \nonumber \\
&&\delta u_{r,R}^n(x)    = \frac{1}{R-r} \int ^{x+R/2-r}_{x-R/2} \delta u_r^n(x') dx', \\
&&v(x+r)v(x)_R           = \frac{1}{R-r} \int ^{x+R/2-r}_{x-R/2} v(x'+r)v(x') dx'. \nonumber
\end{eqnarray}
Here $x$ is the center of the segment, and $r < R$. The mean rate of energy dissipation is $\varepsilon_R = 15 \nu (\partial _x v)^2_R /2$, which yields the Kolmogorov velocity $u_{{\rm K},R} = (\nu \varepsilon _R)^{1/4}$ and the Kolmogorov length $\eta _R =(\nu ^3 / \varepsilon _R)^{1/4}$. We also have the mean total energy, $v^2_R = v(x+r)v(x)_R$ for $r=0$, and the microscale Reynolds number, Re$_{\lambda,R} = 2^{1/2} v^2_R / \nu [(\partial _x v)^2_R]^{1/2}$. The mean rate of energy transfer, however, is not available from our experimental data. Fig. \ref{f1}(a) shows an example of $\delta u_{r,R}^2/u_{{\rm K},R}^2$, which differs from $\langle \delta u_r^2 \rangle / u_{\rm K}^2$.

\subsection{Distribution of fluctuation}
\label{s31}

Over a range of $R$, we study statistics of $\delta u_{r,R}^2/u_{{\rm K},R}^2$ among segments. The scale $r$ is fixed at $10\eta _R$ in the dissipation range and $100\eta _R$ in the inertial range. Since $\delta u_r$ is available only at discrete scales $r$ that are multiples of the sampling interval $U/f_s$, $\delta u_{10\eta _R,R}^2$ and $\delta u_{100\eta _R,R}^2$ are obtained through interpolation by incorporating the fluctuation of $\eta_R$ among segments.

\begin{figure}[b]
\resizebox{8cm}{!}{\includegraphics*[6cm,10.7cm][16cm,27cm]{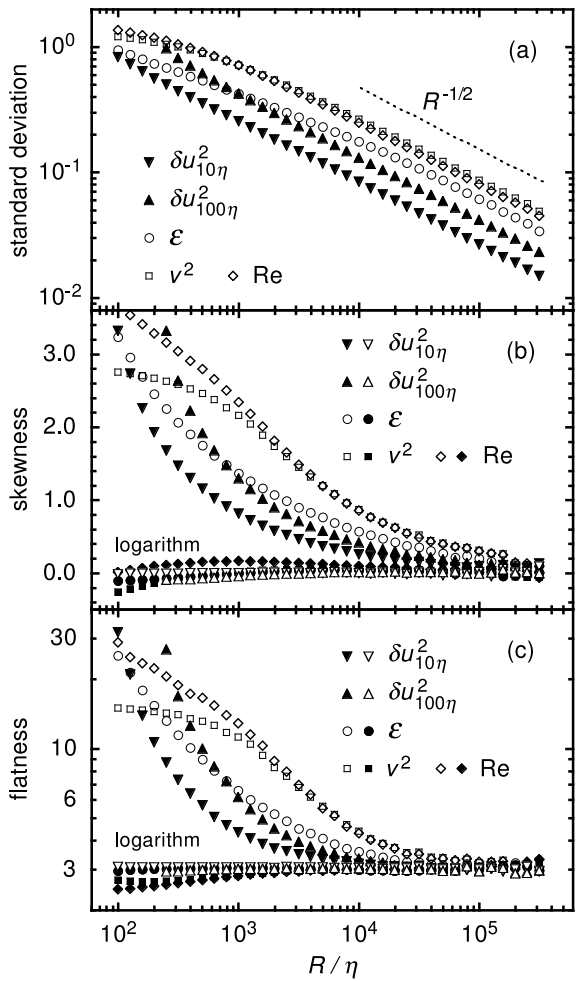}}
\caption{\label{f2} Statistics of $\delta u _{10\eta _R,R}^2/u_{{\rm K},R}^2$ (filled triangles down), $\delta u _{100\eta _R,R}^2/u_{{\rm K},R}^2$ (filled triangles up), $\varepsilon _R$ (open circles), $v^2 _R$ (open squares), and Re$_{\lambda,R}$ (open diamonds) as a function of $R/\eta$. (a) Standard deviation normalized by the average. The dotted line indicates the $R^{-1/2}$ scaling. (b) Skewness factor. (c) Flatness factor. Also shown are the skewness and flatness factors of $\ln (\delta u _{10\eta _R,R}^2/u_{{\rm K},R}^2)$ (open triangle down), $\ln (\delta u _{100\eta _R,R}^2/u_{{\rm K},R}^2)$ (open triangles up), $\ln \varepsilon _R$ (filled circles), $\ln v^2 _R$ (filled squares), and $\ln {\rm Re}_{\lambda,R}$ (filled diamonds).}
\end{figure} 
\begin{figure}[t]
\resizebox{8cm}{!}{\includegraphics*[6cm,15.7cm][16cm,27cm]{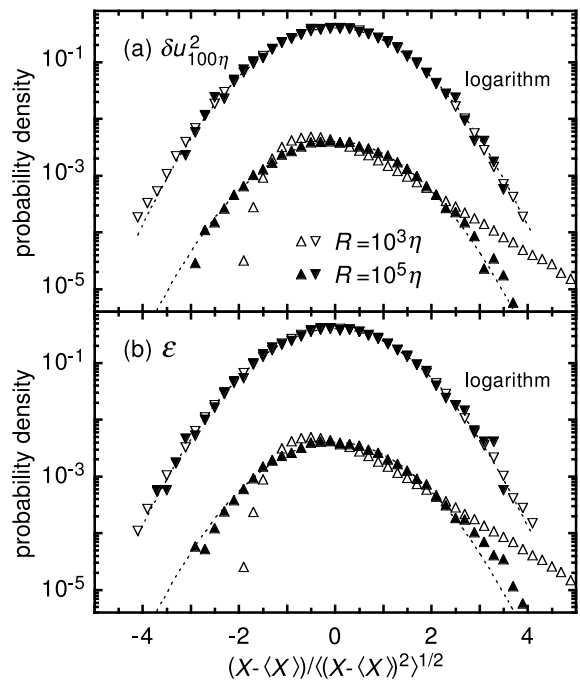}}
\caption{\label{f3} (a) Probability densities of $X_R = \delta u _{100\eta _R,R}^2/u_{{\rm K},R}^2$ (triangles up) and $\ln (\delta u _{100\eta _R,R}^2/u_{{\rm K},R}^2)$ (triangles down) at $R = 10^3 \eta$ (open triangles) and $10^5 \eta$ (filled triangles) as a function of $(X_R - \langle X_R \rangle)/\langle (X_R - \langle X_R \rangle)^2 \rangle^{1/2}$. Those of $X_R = \delta u _{100\eta _R,R}^2/u_{{\rm K},R}^2$ are shifted by a factor of $10^2$. The dotted lines denote the Gaussian distribution. (b) Same as (a) but for $X_R = \varepsilon _R$ (triangles up) and $\ln \varepsilon _R$ (triangles down).}
\end{figure} 

Figure \ref{f2}(a) shows the standard deviation. The fluctuation of $\delta u_{r,R}^2/u_{{\rm K},R}^2$ at a fixed $r/\eta_R$ is significant even when $R$ is large. In individual segments, Kolmogorov's theory\cite{k41} does not hold. The mean rate of energy transfer that determines $\delta u_{r,R}^2$ is not in equilibrium with the mean rate of energy dissipation $\varepsilon_R$ that determines $u_{{\rm K},R}^2$ and $\eta_R$. The degree of this nonequilibrium fluctuates among segments and thereby induces the observed fluctuation of $\delta u_{r,R}^2/u_{{\rm K},R}^2$.

The mean rate of energy transfer also fluctuates among scales in each segment, which does not necessarily have the inertial-range scaling $\delta u_{r,R}^2 \propto r^{2/3}$ [Fig. \ref{f1}(a)].\cite{note2} If each segment had this scaling, the fluctuation of $\delta u_{r,R}^2/u_{{\rm K},R}^2$ among segments at a fixed $r/\eta_R$ in the inertial range would correspond to the fluctuation of the Kolmogorov constant $\delta u_{r,R}^2/(r \varepsilon_R)^{2/3}$.

Figures \ref{f2}(b) and \ref{f2}(c) show the skewness and flatness factors of $\ln (\delta u _{r,R}^2 /u_{{\rm K},R}^2)$ (open triangles). They are close to the Gaussian values of 0 and 3. Thus, at least as a good approximation, the distribution of $\delta u _{r,R}^2 /u_{{\rm K},R}^2$ is lognormal. This is also the case in $\varepsilon _R$ (filled circles) and at $R \gtrsim 10^3 \eta \simeq L_u$ in $v^2_R$ and Re$_{\lambda,R}$ (filled squares and diamonds), while the mean rate of energy transfer should not have a lognormal distribution because it changes its sign. Examples of the probability density functions are shown in Fig. \ref{f3}. The lognormal distribution of $\varepsilon_R$ was discussed as a tentative model by Obukhov.\cite{o62}

A lognormal distribution stems from some multiplicative stochastic process, e.g., a product of many independent stochastic variables with similar  variances. To its logarithm, if not too far from the average, the central limit theorem applies. For the lognormal distributions observed here, the process is related with the energy transfer. While the mean energy transfer is to a smaller scale and is significant between scales in the inertial range alone, the local energy transfer is either to a smaller or larger scale and is significant between all scales.\cite{mouri06,mininni06,ok92} Any scale is thereby affected by itself and by many other scales. They involve large scales because the lognormal distributions are observed up to large $R$. There is no dominant effect from a few specific scales, in order for the central limit theorem to be applicable.

At $R \gtrsim 10^5 \eta \simeq 10^2 L_u$, there are alternative features.\cite{mouri06,kg02} The standard deviations scale with $R^{-1/2}$ [Fig. \ref{f2}(a)]. The skewness and flatness factors of $\delta u _{r,R}^2 /u_{{\rm K},R}^2$, $\varepsilon _R$, $v^2_R$, and Re$_{\lambda,R}$ are close to the Gaussian values [Figs. \ref{f2}(b) and \ref{f2}(c): filled triangles, open circles, open squares, and open diamonds; see also Fig. \ref{f3}]. Their distributions are regarded as Gaussian rather than lognormal, although this has to be confirmed in future using high-order moments or probability densities at the tails for the larger number of segments. The $R^{-1/2}$ scaling and Gaussian distribution are typical of fluctuations in thermodynamics and statistical mechanics,\cite{ll79} for which no correlation is significant at scales of interest as in our case at $r \gtrsim 10^5 \eta$ [Fig. \ref{f1}(b)].

\begin{figure}[t]
\resizebox{8cm}{!}{\includegraphics*[6cm,15.7cm][16cm,27.1cm]{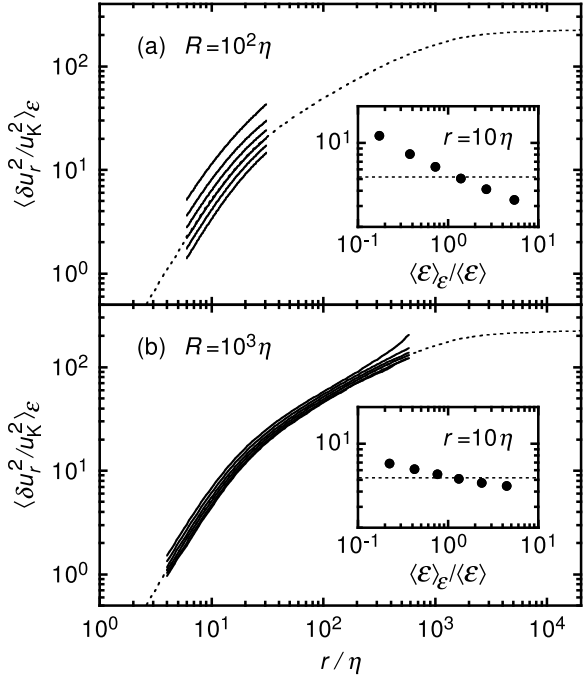}}
\caption{\label{f4} (a) $\langle \delta u_{r,R}^2/u_{{\rm K},R}^2 \rangle _{\varepsilon}$ at $R = 10^2 \eta$ as a function of $r/\eta_R$ (solid lines) and $\langle \delta u_r^2 \rangle/u_{\rm K}^2$ as a function of $r/\eta$ (dotted line). The inset shows the dependence of $\langle \delta u_{10\eta _R,R}^2/u_{{\rm K},R}^2 \rangle _{\varepsilon}$ on $\langle \varepsilon _R \rangle _{\varepsilon}/ \langle \varepsilon \rangle$. The horizontal dotted line indicates the value of $\langle \delta u_{10\eta}^2 \rangle/u_{\rm K}^2$. (b) Same as (a) but at $R = 10^3 \eta$. }
\vspace{0.05cm}
\end{figure} 
\begin{figure}[t]
\resizebox{8cm}{!}{\includegraphics*[6cm,15.7cm][16cm,27.1cm]{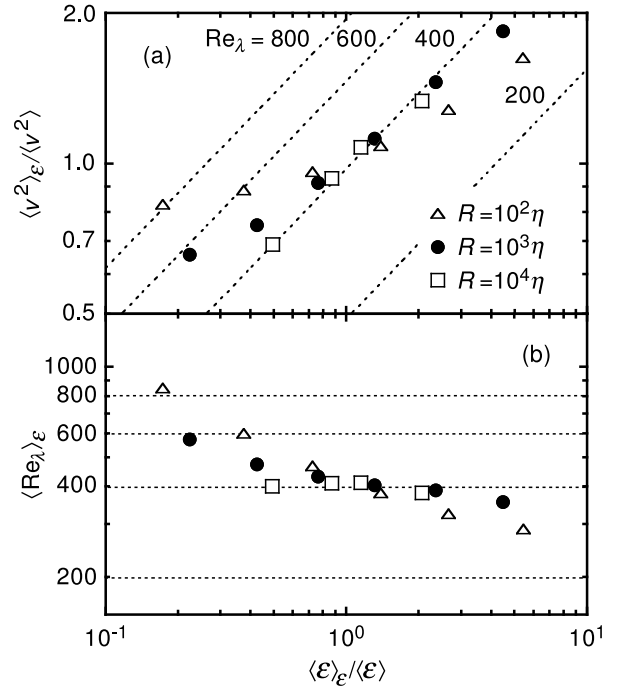}}
\caption{\label{f5} (a) $\langle v_R^2 \rangle _{\varepsilon}/ \langle v^2 \rangle$ as a function of $\langle \varepsilon _R \rangle _{\varepsilon}/ \langle \varepsilon \rangle$ at $R = 10^2\eta$ (triangles), $10^3 \eta$ (circles), and $10^4 \eta$ (squares).  Along the dotted lines, the microscale Reynolds number $15^{1/2} \langle v_R^2 \rangle _{\varepsilon} / (\nu \langle \varepsilon _R \rangle _{\varepsilon})^{1/2}$ is constant. (b) $\langle \mbox{Re}_{\lambda,R} \rangle _{\varepsilon}$ as a function of $\langle \varepsilon _R \rangle _{\varepsilon}/ \langle \varepsilon \rangle$. The symbols are the same as in (a).}
\end{figure} 

\subsection{Correlation between fluctuations}
\label{s32}

The fluctuations among segments at $R \gtrsim L_u$ have interesting correlations. Since these correlations are weak, they are extracted by following Obukhov,\cite{o62} i.e., by averaging over segments with similar $\varepsilon _R$. Specifically, we use conditional averages, denoted by $\langle \cdot \rangle _{\varepsilon}$, for ranges of $\varepsilon _R$ separated at $\langle \varepsilon \rangle /4$, $\langle \varepsilon \rangle /2$, $\langle \varepsilon \rangle$, $2\langle \varepsilon \rangle$, and $4 \langle \varepsilon \rangle$.

Figure \ref{f4} shows $\langle \delta u_{r,R}^2/u_{{\rm K},R}^2 \rangle _{\varepsilon}$ as a function of $r/\eta_R$. When $R = 10^3 \eta \simeq L_u$ [Fig. \ref{f4}(b)], $\langle \delta u_{r,R}^2/u_{{\rm K},R}^2 \rangle _{\varepsilon}$ is independent of $\langle \varepsilon _R \rangle _{\varepsilon}$. The former varies only by a factor of 1.5 while the latter varies by a factor of 20. When $R = 10^2 \eta \simeq 10^{-1} L_u$ [Fig. \ref{f4}(a)], $\langle \delta u_{r,R}^2/u_{{\rm K},R}^2 \rangle _{\varepsilon}$ is not independent of $\langle \varepsilon _R \rangle _{\varepsilon}$.

The implication of the above result is that the mean rate of energy transfer that determines $\delta u_{r,R}^2$ correlates with the mean rate of energy dissipation $\varepsilon_R$ that determines $u_{{\rm K},R}^2$ and $\eta_R$ among segments with $R \gtrsim L_u$. Here $L_u$ is the typical scale for energy-containing eddies. Most of the energy of such an eddy is transferred through scales and dissipated within its own volume. Thus, each energy-containing eddy tends toward equilibrium between the mean rates of energy transfer and dissipation. This tendency does not exist at $R \lesssim L_u$.\cite{note3} Within an energy-containing eddy, the spatial distribution of $\varepsilon$ is not homogeneous. In fact, the $\varepsilon$ correlation is significant at $r \lesssim L_u$ [Fig. \ref{f1}(b)].

Therefore, in order for statistics such as $\delta u_{r,R}^2/u_{{\rm K},R}^2$ and $\langle \delta u_{r,R}^2/u_{{\rm K},R}^2 \rangle _{\varepsilon}$ to have physical meanings, the minimum segment length is about $L_u$. We are to discuss that such segments individually reflect the large-scale flow.

Figure \ref{f5}(a) shows $\langle v^2_R \rangle _{\varepsilon} / \langle v^2 \rangle$ as a function of $\langle \varepsilon _R \rangle _{\varepsilon} / \langle \varepsilon \rangle$. The former varies with the latter. When $R \ge 10^3 \eta \simeq L_u$, the microscale Reynolds number $\propto \langle v^2_R \rangle _{\varepsilon} / \langle \varepsilon _R \rangle ^{1/2}_{\varepsilon}$ is almost constant at the value for the whole record, Re$_{\lambda} = 409$. Thus, segments obey a correlation between $v^2_R$ and $\varepsilon_R$ characterized by the Re$_{\lambda}$ value. Since Re$_{\lambda}$ is determined by the large-scale flow, it follows that the large-scale flow affects each of the segments. A consistent result is obtained for $\langle \mbox{Re}_{\lambda,R} \rangle _{\varepsilon}$ in Fig. \ref{f5}(b).

\begin{figure}[t]
\resizebox{8cm}{!}{\includegraphics*[6cm,20.7cm][16cm,27cm]{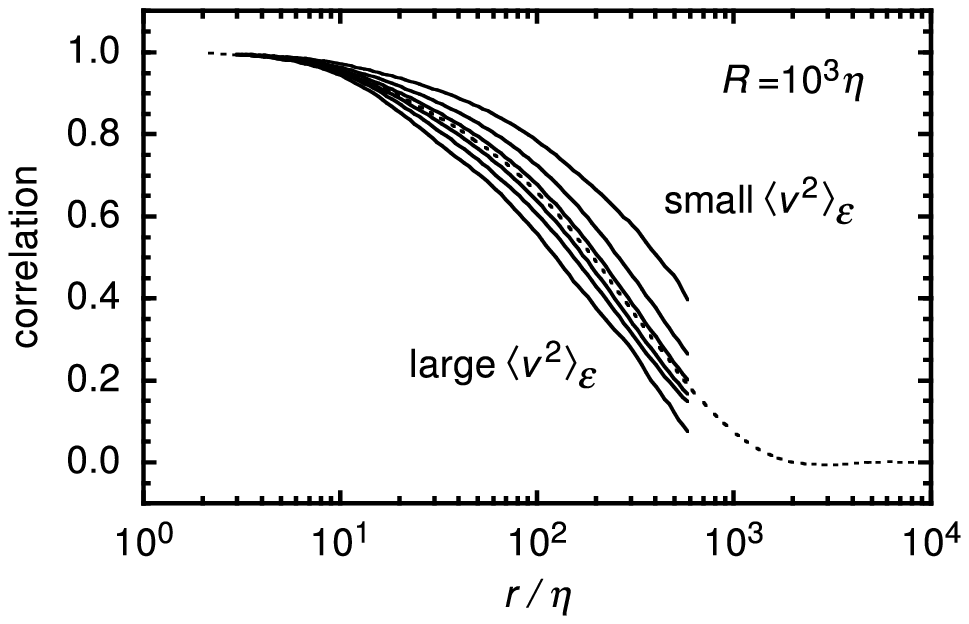}}
\caption{\label{f6} $\langle v(x+r)v(x)_R \rangle _{\varepsilon} / \langle v^2_R \rangle _{\varepsilon}$ at $R = 10^3 \eta$ as a function of $r/\eta_R$ (solid lines) and $\langle v(x+r)v(x) \rangle / \langle v^2 \rangle$ as a function of $r/\eta$ (dotted line).}
\vspace{0.5cm}
\resizebox{8cm}{!}{\includegraphics*[6cm,15.7cm][16cm,27cm]{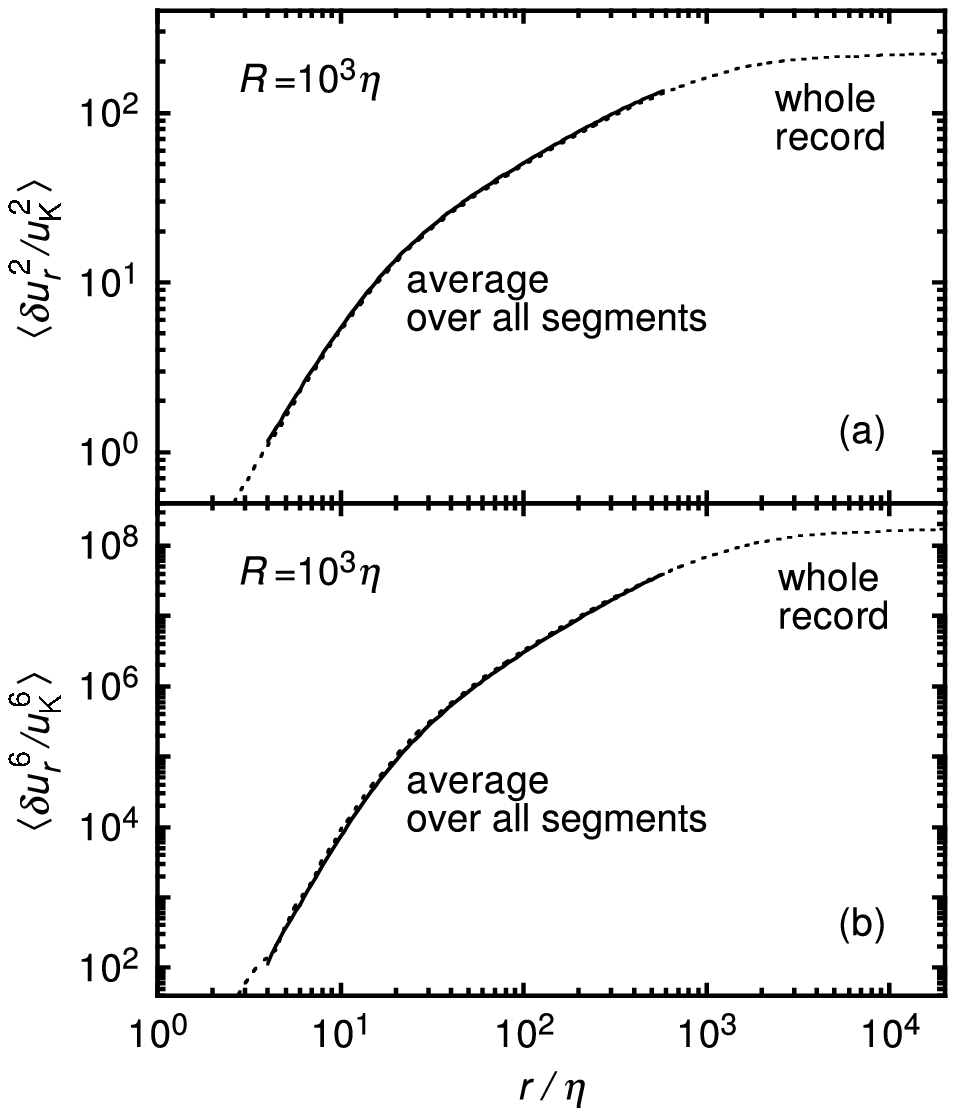}}
\caption{\label{f7} $\langle \delta u_{r,R}^n/u_{{\rm K},R}^n \rangle$ at $R = 10^3 \eta$ as a function of $r/\eta_R$ (solid line) and $\langle \delta u_r^n \rangle/u_{\rm K}^n$ as a function of $r/\eta$ (dotted line). (a) $n = 2$. (b) $n = 6$.}
\end{figure}

The above tendency toward a constant microscale Reynolds number originates in energy-containing scales. In general, through an empirical relation $\langle \varepsilon \rangle \propto \langle v^2 \rangle ^{3/2} /L_v$, Re$_{\lambda}$ is related to the Reynolds number for energy-containing scales as ${\rm Re}_{\lambda} \propto (\langle v^2 \rangle ^{1/2} L_v / \nu)^{1/2}$. Then, Re$_{\lambda}$ is constant if $L_v \propto \langle v^2 \rangle^{-1/2}$. Fig. \ref{f6} shows $\langle v(x+r)v(x)_R \rangle _{\varepsilon} / \langle v^2_R \rangle _{\varepsilon}$, which extends to larger scales for smaller $\langle v_R^2 \rangle _{\varepsilon}$. The correlation length $\int^{\infty}_{0} \langle v(x+r) v(x)_R \rangle _{\varepsilon} dr / \langle v^2_R \rangle _{\varepsilon}$ should be accordingly larger.

The process for each segment to reflect the large-scale flow could be related to the energy transfer. As noted before, the energy transfer couples all scales. The energy transfer itself is affected by the large-scale flow. This is because energy is transferred between two scales via an interaction with some other scale. When the interaction occurs with a large scale, the energy transfer is strong.\cite{mininni06,ok92} 

\subsection{Unimportance of $\varepsilon$ fluctuation}
\label{s33}

Obukhov\cite{o62} discussed that, through the fluctuation of the $\varepsilon_R$ value, the large-scale flow affects small-scale statistics for the whole region such as $\langle \delta u_r^n \rangle / u_{\rm K}^n$. The reason is that $\langle \delta u_r^n \rangle$ is obtained at a fixed $r$, regardless of the fluctuations of $u_{{\rm K},R}^n$ and $\eta_R$ induced by the fluctuation of $\varepsilon_R$. We are to discuss that the fluctuation of the $\varepsilon_R$ value is not important so far as $R \gtrsim L_u$. This condition on $R$ is required for the $\varepsilon_R$ value to correlate with and thus statistically represent the mean rate of energy transfer that determines $\delta u_{r,R}^n$.

Figure \ref{f7} compares $\langle \delta u_r^n \rangle/u_{\rm K}^n$ to $\langle \delta u_{r,R}^n/u_{{\rm K},R}^n \rangle$, i.e., average of $\delta u_{r,R}^n/u_{{\rm K},R}^n$ at each $r/\eta_R$ over all segments, with $R = 10^3 \eta \simeq L_u$. Even for $n = 6$ [Fig. \ref{f7}(b)], they are not distinguishable [see also Fig. \ref{f4}(b)]. Although $\varepsilon _R$ and hence $u_{K,R}$ and $\eta_R$ fluctuate among segments, these fluctuations are not large enough for $\langle \delta u_r^n \rangle/u_{\rm K}^n$ to differ from $\langle \delta u_{r,R}^n/u_{{\rm K},R}^n \rangle$. This conclusion is general. In various flows,\cite{c03,po97,mouri06} including an atmospheric boundary layer at Re$_{\lambda} \simeq 9000$, the standard deviation of $\varepsilon _R/\langle \varepsilon_R \rangle$ at $R \simeq L_u$ is close to the value obtained here [Fig. \ref{f2}(a)]. Hence, through the fluctuation of the $\varepsilon_R$ value, the large-scale flow does not affect small-scale statistics for the whole record. It was suggested that the large-scale flow does affect $\langle \delta u_r^n \rangle/u_{\rm K}^n$, even in the scaling exponents.\cite{k92,sd98,mininni06} If this is the case, the effect is already inherent in the individual segments. They are unlikely to be ``pure\cite{o62}'' or elementary.

\section{Concluding remarks}
\label{s4}

Using segments of a long record of velocity data obtained in grid turbulence, we have studied fluctuations of statistics such as $\delta u_{r,R}^2/u_{{\rm K},R}^2$, $\varepsilon_R$, and $v_R^2$. The fluctuations are significant and have lognormal distributions at least as a good approximation (Figs. \ref{f2} and \ref{f3}). In each segment, the mean rate of energy transfer that determines $\delta u_{r,R}^2$ is not in equilibrium with the mean rate of energy dissipation $\varepsilon_R$ that determines $u_{{\rm K},R}^2$ and $\eta_R$. These two rates still correlate among segments with $R \gtrsim L_u$ (Fig. \ref{f4}), which tend toward equilibrium between the two rates. Also between $\varepsilon_R$ and $v_R^2$, there is a correlation characterized by Re$_{\lambda}$ for the whole record (Fig. \ref{f5}). Thus, the large-scale flow affects each of the segments.

The observed fluctuations depend on $L_u$ and Re$_{\lambda}$, which in turn depend on the configuration for turbulence production, e.g., boundaries such as the grid used in our experiment. Nevertheless, the significance of those fluctuations implies that they have been developed in turbulence itself. Their lognormal distributions are explained by a multiplicative stochastic process in turbulence, which is related with the energy transfer among scales. The correlations among the fluctuations are also explained in terms of the energy transfer.

Previous studies\cite{mouri06,k92,pgkz93,ss96,sd98,mininni06} suggested that the large-scale flow affects some small-scale statistics, although this has to be confirmed at higher Re$_{\lambda}$ where large and small scales are more distinct.\cite{ab06} If the effect really exists, it is inherent individually in the segments.

Our study was motivated by Obukhov's discussion.\cite{o62} It implies the presence of equilibrium between the mean rates of energy transfer and dissipation in an ensemble of segments with similar values of $\varepsilon_R$. This is the case at $R \gtrsim L_u$ (Fig. \ref{f4}).\cite{note4} Also as discussed by Obukhov, the distribution of $\varepsilon_R$ is lognormal at least as a good approximation (Figs. \ref{f2} and \ref{f3}). However, although Obukhov discussed that the large-scale flow affects small-scale statistics through the fluctuation of the $\varepsilon_R$ value, this is not the case (Fig. \ref{f7}).

The lognormal distributions observed here have to be distinguished from those proposed by Kolmogorov.\cite{k62} While he was interested in small-scale intermittency and studied $\varepsilon_r$ and $\delta u_r^n$ at small $r$ to obtain their scaling laws, we are interested in large-scale fluctuations and have studied $\varepsilon_R$ and $\delta u_{r,R}^n$ at small $r$ but at large $R$. The scaling laws of $\varepsilon_R$ and $\delta u_{r,R}^n$ are not necessary. In addition, the lognormality has been attributed to a different process. Hence, our study is not necessarily concerned with the well-known problems of Kolmogorov's lognormal model, e.g., violation of Novikov's inequality\cite{n71} for scaling exponents. Still exists a possibility that small-scale intermittency is affected by large-scale fluctuations. The study of this possibility is desirable.

There were no studies of statistics among segments with large $R$. Hence, we have focused on grid turbulence, which is simple and thus serves as a standard. For flows other than grid turbulence, the fluctuations of statistics among segments are expected to be significant as well. In fact, regardless of the flow configuration and the Reynolds number, the large-scale fluctuation of $\varepsilon_R$ is significant.\cite{po97,c03,mouri06}   Those fluctuations are also expected to have lognormal distributions and mutual correlations as observed here because they are due to the energy transfer in turbulence itself. However, grid turbulence is free from shear. It was previously found that $\delta u_r$ correlates with $u$ in shear flows such as a boundary layer but not in shear-free flows.\cite{pgkz93,ss96,sd98} The fluctuations of statistics among segments might be somewhat different in a shear flow. To this and other flow configurations, it is desirable to apply our approach.

\begin{acknowledgments}
We are grateful to K. R. Sreenivasan for inspiring this study and for helpful comments and also to M. Tanahashi for helpful comments.
\end{acknowledgments}



\begin{thebibliography}{999}
\bibitem{note0} If we had obtained snapshots of the velocity field in the wind tunnel, we should have observed various eddies. Their strengths should have been random but on average decayed along the streamwise direction. Over scales along the streamwise direction, statistics of velocity fluctuations should have suffered from the decay, which is not relevant to our study. Over scales along the spanwise direction, so far as they were not too large, statistics of velocity fluctuations should have been identical to those for $u(x)$ and $v(x)$.
\bibitem{note1} The alternative interpretation is that such fluctuations correspond to spatial fluctuations of some virtual turbulence. Local values of $u(x)$ and $v(x)$ represent local regions of actual turbulence, and they are continuously connected up to largest scales. Throughout the scales, $u(x)$ and $v(x)$ obtained here are consistent with those measured at a certain time in spatially homogeneous but temporally decaying turbulence (see Appendix of Ref. \onlinecite{mouri06}).
\bibitem{note2} The reader might consider that $\delta u_{r,R}^2 \propto r^{2/3}$ is absent in some segments because their Re$_{\lambda,R}$ values are not high enough. This is not the case. For example, the segment shown in Fig. \ref{f1}(a) has Re$_{\lambda,R} \simeq 1000$.
\bibitem{note3} For such a segment, statistics do not strictly have local isotropy, and hence $\varepsilon_R = 15 \nu (\partial _x v)^2_R /2$ is not strictly exact. The associated uncertainty is nevertheless small and not serious to our study. In fact, among segments with $R = 10^2 \eta$ and $10^3 \eta$, respectively, the correlation coefficients are as high as 0.89 and 0.95 between $15 \nu (\partial _x v)^2_R /2$ and $15 \nu [(\partial _x u)^2_R+(\partial _x v)^2_R/2]/2$. The value of the latter tends to be more close to the true $\varepsilon_R$ value (see Ref.  \onlinecite{c03}), although it is not used in our study because of possible spurious correlations with the values of $\delta u_r$ over small $r$.
\bibitem{note4} To examine Obukhov's discussion in a strict manner, we have to use $\langle \delta u_{r,R}^2 \rangle _{\varepsilon}/(\nu \langle \varepsilon_R \rangle_{\varepsilon})^{1/2}$ instead of $\langle \delta u_{r,R}^2/u_{{\rm K},R}^2 \rangle _{\varepsilon}$. Even in this case, the result is almost the same as in Fig. \ref{f4}. An analogous situation is seen in Fig. \ref{f5}, where the Reynolds number obtained from $\langle v_R^2 \rangle _{\varepsilon}$ and $\langle \varepsilon_R \rangle_{\varepsilon}$ is almost the same as $\langle \mbox{Re}_{\lambda,R} \rangle _{\varepsilon}$.
\end{thebibliography}
\end{document}